\documentstyle[preprint,eqsecnum,aps,prb]{revtex}

\begin{document}
%\draft
%
\title{Quantum Computation Based on Magic-Angle-Spinning Solid State Nuclear Magnetic Resonance
Spectroscopy}

\author{Shangwu Ding$^{1,2,4}$,
Charles A. McDowell$^{3}$,
Chaohui Ye$^{1,2}$, Mingsheng Zhan$^{1,2}$, Xiwen Zhu$^{1,2}$,
Kelin Gao$^{1,2}$, Xianping Sun$^{1,2}$,
Xi-An Mao$^{1,2}$, Maili Liu $^{1,2}$}
\address{$^{1}$National Laboratory of Magnetic Resonance
and Atomic and Molecular Physics,  P. O. Box 71010, Wuhan, Hubei 430071, China}
\address{$^{2}$Wuhan Institute of Physics and Mathematics, The Chinese
Academy of Sciences, P. O. Box 71010, Wuhan, Hubei 430071, China}
\address{$^{3}$Department of Chemistry, University of British Columbia,
2036 Main Mall, Vancouver, British Columbia, Canada, V6T 1Z1}
\address{$^{4}$Department of Chemistry, National 
Sun Yat-Sen University, Kaohsiung, Taiwan 807, Republic of China}
\date{\today}
\maketitle
\begin{abstract}
Magic-angle spinning (MAS) solid state nuclear magnetic resonance (NMR) spectroscopy 
is shown to be a promising technique for implementing quantum computing.
The theory underlying the principles of quantum computing
with nuclear spin systems undergoing MAS is formulated in the framework of 
formalized quantum Floquet theory. The procedures for realizing state labeling, state transformation and coherence selection in Floquet space are given.
It suggests that by this method, the largest 
number of qubits can easily surpass that achievable with other
techniques. Unlike other modalities proposed for quantum 
computing, this method enables one to adjust the dimension
of the working state space, meaning the number of qubits can be readily
varied. The universality of quantum computing in Floquet space with solid state NMR is discussed and a demonstrative experimental implementation of Grover's search is given.
\end{abstract}
\widetext
\section{introduction}
%\paragraph{}
As early as the late 1950's, 
Landauer and Bennet {\it et al.}\cite{land61,benn73,benn82,beni82,fred82,benn89},
investigated the
effects of physical laws on computing, such as the
reversibility of a computing operation and the minimal energy
required to transmit a bit of information.
Feynman\cite{feyn82}, on the other hand, was studying the
fundamental limitations of
quantum mechanics on the capacity of (classical) computers.
The most important question in these works was what 
would it happen if computing logic is not presumably given but rather 
determined by physical laws, particularly, quantum mechanical laws?
With the rapid development of very large scale integrated circuitry
technology, above question seemed to become important in the early
1980's;
that can be rephrased as, what would it happen if the chip size were
made 
so small that one chip contains very few, even just one impurity electron.
That background of scientific development initiated quantum computing
research. However, quantum computing 
was basically dormant in the decade of the 1980's. 
It has since gained increasing attention once the power of a
hypothetical
quantum computer was revealed,
particularly, through the works of Deutsch\cite{deut85,deut89,deut92,bert94,deut95}  
Shor\cite{shor94,gald96},
Lloyd\cite{lloy95} etc. 
Deutsch\cite{deut85,deut89,deut95} showed that genuine 
and massive parallelism can be achieved. 
Lloyd\cite{lloy95} proposed a quantum computing prototype
that has subsequently been followed.  
Shor\cite{shor94} demonstrated the power of quantum computer 
in solving the famous and all-important problem in number theory
and public key cryptography system,
{\it i.e.} the prime factoring of large integers.
Shor {\it et al.}\cite{shor95,gald96,divi96}, Gottesman\cite{gott96,gott98},
Steane\cite{stea96a,stea96b,stea99},
Schumacher\cite{schu95}and Preskill\cite{pres98} and others 
invented a variety of quantum error correction schemes 
that are crucial to the realization of long-time quantum computing.\\
%\paragraph{}
Since then, theoretical publications have appeared 
with increasing frequency, encompassing
almost every aspect of computing theory(for review, see, {\it e.g.}, \cite{bere96,stea98}).
Remarkable progress in experimental implementation and model proposals
also has been made in utilizing an extensive 
repertoire of sophisticated experimental techniques 
including atomic interferometry\cite{bare95},
quantum electrodynamic cavity\cite{turc95,slea95,domo95,pell95,monr95},
ion trap\cite{cira95,jeff95}, polarized photons\cite{torm96},
nuclear spins embedded in an electron system in the quantum Hall
regime\cite{priv97},
quantum dots\cite{bona98}, Josephson junction\cite{mooi99},
electrons in liquid helium\cite{plat99},
nuclear spins in doped silicon devices\cite{kane98},
single Cooper pair\cite{naka99}, Rydberg atom\cite{ahn00}
and liquid NMR
\cite{gers97a,cory97,gers97b,chua98a,chua98b,jone98a,jone98b,cory98,chua98c,lind98,lind99,zhu99,madi98,marj00,marx00,cory99,fung00}. 
Differing from other techniques, the NMR prototype
uses bulk samples hence an ensemble of nuclear spins rather than pure
quantum
mechanical systems. 
Among the above experimental prototypes,  
NMR is certainly the most promising, to
date: all above methods
except NMR can only simulate a single quantum logic gate such as 
controlled NOT gate,
but NMR can do much more than that, e.g., it can simulate a quantum
network such as the performing of simple arithmetic operations, and a
quantum
computer that can execute simple quantum
algorithms\cite{chua98a,chua98b,chua98c,jone98a,jone98b,zhu99,marx00}. 
The NMR method offers the first realizable quantum computer
operating with more than two qubits, thus providing for the first
time a quantum computer with error correction capacity\cite{cory99}.
All these demonstrations used liquid state NMR
spectroscopy because of its natural high resolution.
While the progress has been remarkable,
one severe difficulty with the NMR quantum computer
is the exponential loss of the signal sensitivity with the 
increase of spin numbers
(hence usable qubits) in the working molecule. It is clear that
establishing
a NMR quantum computer with a capacity of over ten qubits is rather 
challenging\cite{warr97}, if not
impossible even 
though a host of sensitivity-enhancing techniques are 
available\cite{gers97b}. \\
%\paragraph{}
In this paper, we present an alternative, probably more
advantageous, method for performing NMR quantum computing, that is, quantum 
computing based
on solid state NMR involving rotating samples at an angle of
54.74$^{\circ }$, the magic-angle to the applied magnetic field. 
This so-called magic-angle spinning
(MAS)\cite{haeb76,mehr83} NMR can be well formulated using Floquet
theory
\cite{shir65,mari82,zur83,vega84,schm87,kubo90,schm92,naka92,ding98a,ding98b}.
From the point of view of quantum computing, the
Floquet description offers a method to augment the state space,
almost infinitely. In practice, nevertheless, the size of the space is
restricted by the signal sensitivity. However, as shown in our
theoretical analysis elaborated below, this size
can be easily made much larger than that realizable in liquid NMR studies.
For quadrupolar nuclei, the sidebands produced by the rotating
polycrystalline samples can
be as many as thousands or even 
more\cite{jako89,skib91},
meaning usable qubits can be easily achieved, even in excess of
10  
merely by using conventional
NMR techniques, although in this case, the manipulation of quantum states and coherences
is more complicated than for spin-1/2 systems.
The other obvious advantage of solid state NMR is that the
number of spins in a sample is usually much larger than
in a liquid sample of the same size, meaning a significant sensitivity gain.
%\paragraph{}
Our paper is arranged as follows. 
In the Section II, the theory underlying  quantum computing with solid 
state MAS NMR is described; this is the foundation of the
subsequent
sections 
of paper
and future work. Particularly important are the definition of 
the pseudo pure state in Floquet space and its connection with quantum computing.
While the theoretical framework applies to nuclear spin-1/2 systems as
well as to 
quadrupolar nuclear spins, the remainder of the paper will focus on
spin-1/2 systems with chemical shift interactions. 
Section III establishes the correspondence between a pseudo pure state 
and its spectral representation. This is essential to the read-out
function in
NMR quantum computing because the directly detectable signals in NMR arise
from 
the single quantum coherences. The theoretical derivation of the spectral
signal
is demonstrated in Appendix A.
The preparation of pseudo pure state is crucial to quantum computing and
this is discussed in Section IV. Three different methods are
considered. Section IV analyzes the universality of MAS solid state NMR quantum computing. In Section VI is demonstrated the implementation of an important quantum computing algorithm, namely, Grover's search, on a solid state NMR quantum computer. 
The major points of this paper are summarized in the final Section. 
\section{Quantum computing in Floquet space}
\subsection{Quantum Floquet theory of solid state NMR}
%\paragraph{}
A periodic time-dependent Hamiltonian 
such as that for a nuclear spin system in a polycrystalline sample
undergoing rotation at the magic-angle in a static magnetic field
is best described employing Floquet theory. Here we summarize the
well-developed theory from the
perspective of MAS NMR and its significance to quantum computing.
Most generally, the evolution of the
density matrix, $\rho(t)$, of a spin system 
can be written as
\begin{equation} 
\setcounter{equation}{1}
\rho(t)={\hat{T}}e^{-i\int_{0}^{t}dt'H(t')}\rho(0)e^{i\int_{0}^{t}dt'H(t')}.
\end{equation}
It follows, therefore, that evaluating the evolution operator
$U(t)=e^{-i\int_{0}^{t}dt'H(t')}$ is a
central part of spectral lineshape calculations. The straight-forward
procedure is
to use the multi-step method which divides the time interval, $(0, t_{c})$,
where $t_{c}$
is the period of the Hamiltonian, into $N$ equal steps and then one calculates each
step by approximating its Hamiltonian as being time-independent
\begin{equation}
\setcounter{equation}{2}
U(t)=e^{-iH(t_{n})\Delta t}...e^{-iH(t_{i})\Delta t}...e^{-iH(0)\Delta t},
\end{equation}
where $t=nt_{c}/N$.
This usually involves the diagonalization of each instantaneous
Hamiltonian $H(t_{i})$. 
Floquet formalism\cite{zur83,vega84,schm87,kubo90,schm92,naka92,ding98a,ding98b}, on
the other hand, focuses the calculation of the evolution operator
on computing the Floquet Hamiltonian $H_{F}$ by introducing Floquet
states $|rn>$
where $r$ is the state index of $H$ and $n$ is the mode index\cite{zur83,vega84}
\begin{equation}
\setcounter{equation}{3}
<pm|H_{F}|qn>=h_{pq}^{m-n}+n\omega_{c}\delta_{pq}\delta_{mn}
\end{equation}
where $h_{pq}^{k}$ are the Fourier components of the time-dependent
Hamiltonian \begin{equation}
\setcounter{equation}{4}
H_{pq}=\sum_{k}h_{pq}^{k}e^{ik\omega_{c}t}
\end{equation}
The evolution operator then can be calculated from the following
expression
\begin{eqnarray}
\setcounter{equation}{5}
U_{pq}(t)&=&\sum_{n=-\infty}^{\infty}<pn|e^{-iH_{F}t}|q0>
e^{in\omega_{c}t}, \nonumber \\
&=&\sum_{r}\sum_{n,k=-\infty}^{
\infty}<pn|\lambda_{r}^{0}><\lambda_{r}^{0}|qk>e^{-i(q_{r}-n\omega_{c})t}.
\end{eqnarray}
where the index $r$ runs over the Hilbert space defined by $H$.
For the sake of generality, the Hamiltonian is assumed to be anisotropic,
{\it i.e.} $H\equiv H(\alpha, \beta, \gamma, t)$ where
$\alpha, \beta$ and
$\gamma$ are the Euler angles describing the interaction tensors 
relative to the laboratory frame (they can be set to zero for solution NMR
cases).
$|\lambda_{r}^{n}>$ and $\lambda_{r}^{n}$ are the eigenstate and
eigenvalues of the
Floquet Hamiltonian, respectively
\begin{eqnarray}
\setcounter{equation}{6}
H_{F}|\lambda_{r}^{n}>&=&\lambda_{r}^{n}|\lambda_{r}^{n}> \\
q_{r}&=&\lambda_{r}^{n}-n\omega_{c}
\end{eqnarray}
Eq.(5) can be rewritten as\cite{mari82,zur83,vega84}
\begin{equation}
\setcounter{equation}{8}
U(t)=P(t)e^{-iQt}P(0)^{-1}
\end{equation}
where $Q$ is traceless and diagonal with diagonal elements $q_{r}$,
and $P(t)$ the Floquet amplitude is defined as
\begin{equation}   
\setcounter{equation}{9}
P_{pq}^{n}=<pn|\lambda_{q}^{0}>
\end{equation}
At first sight, the estimated magnitude of $U(t)$ can be made
almost exact because the
values of ${t}$ can be chosen in arbitrarily small increments. However, in
most realistic cases, the Floquet
Hamiltonian $H_{F}$ cannot be solved exactly: it requires the use of a
perturbation expansion, or equivalently, matrix diagonalization.
As has been shown, the order of the expansion series, or
the chosen dimension of $H_{F}$ presents a bound to the accuracy of $U(t)$.

If we further define density matrices, observable operators and evolution
operators in the Floquet basis, a formalized Floquet theory\cite{leva95} 
can be formulated. Specifically, we define 
\begin{equation}
U^{F}(t)=\Sigma_{n,m}U_{n-m}(t)|n><m|e^{-in\omega_{r}t}
\end{equation}
where $U_{n}(t)$ are given by 
\begin{equation}
U(t,t_{0})=\Sigma_{n}U_{n}(t)e^{-in\omega_{r}t_{0}}
\end{equation}
Then the density matrix can be found as
\begin{eqnarray}
\sigma(t)&=&\Sigma_{n,m}<n|\sigma^{F}(t)|m>e^{-i(n-m)\omega_{r}t} \\ \nonumber
&=&\Sigma_{n,m}<n|U^{F}\sigma^{F}(0)U^{F^{-1}}|m>e^{-i(n-m)\omega_{r}t} 
\end{eqnarray}
where $\sigma^{F}$ satisfies the Liouville equation
\begin{equation}
\frac{d\sigma^{F}(t)}{dt}=-i[H^{F},\sigma^{F}(t)]
\end{equation}
and the initial density matrix at time $t=0$ is given by
\begin{equation}
<n|\sigma^{F}|m>=\delta_{n,m}\sigma(0)
\end{equation}
The observable operator is defined as
\begin{equation}
A^{F}=\Sigma_{n,m}A_{n-m}|n><m|=\Sigma_{n,m}A_{m}|n><n-m|
\end{equation}
where $A_{n}$ are the Fourier components of $A(t)$:
\begin{equation}
A(t)=\Sigma_{n}A_{n}e^{-in\omega_{r}t}
\end{equation}
Note that definition of the Floquet Hamiltonian $H^{F}$ in the
original work\cite{leva95} is different from Eq.(3).
The formalized form of the detection observable $D$ then is easily found from 
the requirement
\begin{equation}
S(t)={\rm Tr}[D\sigma (t)]={\rm Tr}[\tilde{D}^{F}\sigma^{F}(t)]
\end{equation} 
to be 
\begin{equation}
\tilde{D}^{F}=\Sigma_{n,m}D|n><m|e^{-i(n-m)\omega_{r}t}
=\Sigma_{n,m}D|n><n+m|e^{-im\omega_{r}t}
\end{equation}

Because of its unified and has a concise form, the formalized Floquet theory 
will be used throughout the work.

\subsection{Floquet pseudo pure state}

With the above formalized Floquet theory, the pseudo pure state or 
effective pseudo pure state
introduced by Cory {\it et al.}\cite{cory97}
can be extended to Floquet space as follows:
\begin{equation}
\Psi^{F}=\frac{(1-\alpha){\bf \hat{I}}+\alpha |\phi^{F}><\phi^{F}|}{2^{nk}}
\;\;\;(|\alpha|\le 1) 
\end{equation}
where $n$ is the number of spin-1/2 nuclei, 
$k$ is the (effective) dimension of the "mode" space
and ${\bf \hat{I}}$ is the identity spinor whose matrix form 
is an $nk\times nk$ identity square matrix.
It is easy to verify that the above definition of the 
pseudo pure state satisfies the three criteria given by Cory {\it et al.}\cite{cory97}:
{\it i.e.}, $\Psi^{F}$ defines a pure density matrix 
and vice versa; $\Psi^{F}$ evolves according to the same unitary transform
governing the evolution of a pure density matrix and 
the measurement value of an observable operator over $\Psi^{F}$
and that of the same  observable operator over a pure density matrix
differ in a trivial constant only. Therefore, the pseudo pure state in Floquet space 
can be used to "emulate" quantum computing. 

To specify the Floquet space, in the following sections we will focus on 
the chemical shift interaction. The Euler angles system is defined 
as follows:
the principal axis direction of the chemical shift tensor in the rotor
systems is determined with ($\alpha, \beta, \gamma$)
while the rotor system is specified by ($\omega_{r}t, \theta, 0$)
relative to the laboratory frame, where $\omega_{r}$ is the sample spinning speed.
The most interesting case in solid state NMR as well in 
this work is when the sample spins at the "magic angle", {\it i.e.}
the spinning axis is tilted $\beta=\beta_{m}=54.74^{\circ}$ 
with respect to the static magnetic field, which is called magic-angle-spinning(MAS).  
The chemical shift interaction Hamiltonian then can be written as\cite{haeb76,mehr83}  
\begin{equation}
H_{CS}=-I_{z}\{\delta_{0}+\delta P_{2}(\rm{cos}\theta)[P_{2}(\rm{cos}\beta)
-\frac{\eta}{2}{\rm sin}^{2}\beta{\rm cos}2\gamma]+\frac{\sqrt{3}}{2}\delta\xi (t)\}
\end{equation}
where $\delta_{0}$ is the isotropic chemical shift plus the RF offset,
$\delta, \eta$ are the anisotropy and asymmetry parameters of the chemical shift tensor,
respectively. Denote the three principal values of the chemical shift tensor 
as $\sigma_{11},\sigma_{22},\sigma_{33}$ and assume the convention 
$\sigma_{11}\ge \sigma_{22}\ge \sigma_{33}$, then there are relations 
$\delta_{0}=\frac{1}{3}(\sigma_{11}+\sigma_{22}+\sigma_{33}),
\delta=\sigma_{0}-\sigma_{33}, 
\eta=(\sigma_{11}-\sigma_{22})/\delta$. 
It is noteworthy that at the magic angle, all the anisotropic terms in Eq.(20)
disappear, meaning registered spectra are free from line broadening caused 
by chemical shift interactions. 
This is the most important principle in high resolution solid state NMR.
The time-dependent term $\xi (t)$ in Eq.(20) is given by
\begin{equation}
\xi (t)=C_{1}{\rm cos}(\omega_{r}t)+
S_{1}{\rm sin}(\omega_{r}t)+C_{2}{\rm cos}(2\omega_{r}t)+S_{2}{\rm sin}(2\omega_{r}t)
\end{equation}
where 
\begin{eqnarray} 
C_{1}&=&\frac{1}{2}{\rm sin}2\theta {\rm sin}\beta [-{\rm cos}\beta (\eta {\rm cos}2\gamma +3){\rm
cos}\alpha +\eta{\rm sin}2\gamma{\rm sin}\alpha] \\ \nonumber
S_{1}&=&\frac{1}{2}{\rm sin}2\theta {\rm sin}\beta [{\rm cos}\beta (\eta {\rm cos}2\gamma +3){\rm 
sin}\alpha +\eta{\rm sin}2\gamma{\rm cos}\alpha] \\ \nonumber
C_{2}&=&\frac{1}{2}{\rm sin}^{2}\theta \{[\frac{3}{2}{\rm sin}^{2}\beta -\frac{\eta}{2} {\rm
cos}2\gamma (1+{\rm cos}^{2}\beta)]{\rm cos}2\alpha 
+\eta {\rm cos}\beta {\rm sin}2\gamma {\rm sin}2\alpha \} \\ \nonumber
S_{2}&=&\frac{1}{2}{\rm sin}^{2}\theta \{[-\frac{3}{2}{\rm sin}^{2}\beta -\frac{\eta}{2} {\rm cos}2\gamma
(1+{\rm cos}^{2}\beta)]{\rm sin}2\alpha +\eta {\rm cos}\beta {\rm sin}2\gamma {\rm cos}2\alpha \} 
\end{eqnarray}
From Eqs.(20,21), the Floquet Hamiltonian $H_{CS}^{F}$ can readily 
be found from Eqs.(3,4)\cite{schm92}. 
\begin{eqnarray}
<0n|H^{F}_{CS}|0n>&=& n\omega_{r}-\frac{1}{2}\delta\omega 
<1n|H^{F}_{CS}|1n>=n\omega_{r}+\frac{1}{2}\delta\omega  \\ \nonumber
<0n|H^{F}_{CS}|0n\pm 1>&=&-\frac{\sqrt{3}}{4}\delta_{CS}C_{1} 
<1n|H^{F}_{CS}|1n\pm 1>=\frac{\sqrt{3}}{4}\delta_{CS}C_{1} \\ \nonumber
<0n|H^{F}_{CS}|0n\pm 2>&=&-\frac{\sqrt{3}}{4}\delta_{CS}C_{2} 
<1n|H^{F}_{CS}|1n\pm 2>=\frac{\sqrt{3}}{4}\delta_{CS}C_{2} 
\end{eqnarray}
All the other elements are zero. 
 The chemical shift interaction is a typical "inhomogeneous" interaction,
{\it i.e.},
it's Hamiltonian at different times is always commutable rendering  
it unnecessary to perform the time ordering operation in the
calculation of the unitary evolution operator.  
This is a very important property that helps one 
analyse the evolution of (pseudo pure) quantum states
and simplify the design of quantum computing gates. 
Other important inhomogeneous interactions 
include the electric quadrupolar interaction and heteronuclear 
dipolar coupling which will be discussed in future work.

From above paragraphs, some important implications of the
applications of 
Floquet space 
and solid state NMR to quantum computing in Floquet formalism are
summerized as follows:
First, Floquet space is dimensionally adjustable, {\i.e.}, changing the 
sample spinning
speed $\omega_{r}$ can augment or reduce the effective dimension of
the space meaningful for quantum computing. Moreover, contrary to a usual
quantum mechanical
system, its dimension is not necessarily a power of 2.
Second, in solid state NMR quantum computing, the Hamiltonian hence the 
"operation" can be controlled both with the RF field and sample spinning
speed, which provides
more flexibility than solution NMR quantum computing. 
Third, opposite to solution NMR quantum computing which is not
satisfactory at 
low temperatures (below the melting point of the sample used), solid state
NMR 
is usually more sensitive at temperatures as low as possible.

\section{Spectral representation of Floquet state}
\subsection{Signal readout}
One of the most essential functions of computing is that the output 
can be read out.  
In this section, we give an operational procedure 
on how to "read out" a pseudo pure state of a solid state NMR quantum 
computer. In line with standard NMR spectroscopy which
measures the induction voltage caused by the transverse magnetization vector
of the spin ensemble, this is done 
by observing the spin ensemble (which is in a pseudo pure state).
There are numerous "read" pulses available, but for simplicity and without
losing generality, we use single 90$^{\circ}$ pulse in this work
(This is sufficient for chemical shift interaction but for quadrupolar
interaction an effective observation may demand more complicated 
pulses, which will be discussed in the future). 

The energy levels of a spin-1/2 system in spinning solid NMR 
are labeled as shown in Fig.1(a). The first index
is spin angular momentum quantum number and the second one the
mode. The readout function of an output state is given in Fig.1(b).
Therefore, the FID (free induction decay) signal of 
a pseudo pure state $|pm>$ can be given as 
\begin{equation}
S^{pm}_{+}(t)={\rm
Tr}[{\tilde{I}}_{+}^{F}U_{CS}^{F}U_{90}^{F}\sigma^{Fpm}{U_{90}^{F}}^{-1}
{U_{CS}^{F}}^{-1}]
\end{equation}
where $U_{CS}^{F}$ and $U_{90}^{F}$ are the Floquet evolution operators
corresponding to chemical shift interaction 
and the 90$^{\circ}$ pulse, respectively.
$\sigma^{Fpm}\equiv |pm><pm|$ is a pure state in Floquet space.
${\tilde{I}}_{+}^{F}$ is the observable operator defined by 
\begin{equation}
{\tilde{I}}_{\alpha}^{F}=\Sigma_{m,n}|n>I_{\alpha}<n+m|e^{im\omega_{r}t}\;\;\;\;\alpha=x,y,\pm, 
{\rm etc.} 
\end{equation}
\subsection{The expressions for $U_{90}^{F}$ and $U_{CS}^{F}$}
We assume the 90$^{\circ}$ pulse is along the -x direction in the rotating frame,
the RF Hamiltonian is written as 
\begin{equation}
H_{rf}=\omega_{1}I_{x}\equiv \frac{\omega_{1}}{2}\sigma_{x}
\end{equation}
where $\omega_{1}$ is the RF field strength and $\sigma_{x}$ is the Pauli matrix.
Above equation leads to the following Floquet Hamiltonian
\begin{eqnarray}
H_{rf}^{F}&=&\omega_{1}\Sigma_{n}|n>I_{x}<n|+n\omega_{r}{\bf 1}|n><n| \\ 
\nonumber
&=&\omega_{1}\frac{1}{2}\sigma_{x}\bigoplus \frac{1}{2}\sigma_{x}\bigoplus ...
\frac{1}{2}\sigma_{x}+n\omega_{r}{\bf 1}|n><n|
\end{eqnarray}
The general expression for the evolution operator 
of the RF interaction is then given by 
\begin{equation}
U_{rf}^{F}(t_{p})=e^{-iH_{rf}^{F}t_{p}}=e^{-i[\omega_{1}\Sigma_{n}|n>I_{x}<n|
+n\omega_{r}{\bf 1}|n><n|]t_{p}}
\end{equation}
where $t_{p}$ is the pulse width. In explicit matrix form, Eq.(28) is
\begin{equation}
U_{rf}^{F}(t_{p})=\left(
\begin{array}{lllcccrrr}
.& & & & & & & &  \\
&.& & & & & & & \\
& &.& & & & & & \\
& & & U_{I_{x}} e^{-i\omega_{r}t_{p}}& & & & & \\
& & & & U_{I_{x}}& & & & \\
& & & & &U_{I_{x}} e^{i\omega_{r}t_{p}} & & & \\
& & & & & & .& & \\
& & & & & & & .& \\
& & & & & & & & . 
\end{array} \right)\;
\end{equation}
If the effective dimension of the mode space is $K$ and
the condition $K\omega_{r}t_{p}\rightarrow 0$ is satisfied,
above equation is reduced to 
\begin{equation}
U_{rf}^{F}(t_{p})=\left(
\begin{array}{lllcccrrr}
. & & & & & & & & \\
  &. & & & & & & &\\
 & & .& & & & & & \\
& & & U_{I_{x}} & & & & & \\
& & & & U_{I_{x}}& & & & \\
& & & & &U_{I_{x}} & & & \\
& & & & & & .& & \\
& & & & & & & .& \\
& & & & & & & & . 
\end{array} \right)
\end{equation}
which is a useful simplified expression. 
%\paragraph{}
The explicit expression of the evolution operator of 
chemical shift interaction Hamiltonian can be found from Eqs.(20-22).
Specifically, for a spin-1/2 system, from the Hamiltonian Eq.(20), 
we have
\begin{eqnarray}
U_{CS}(t,t_{0})&=&e^{-i\int_{t_{0}}^{t} dt(\omega_{CS}+\delta_{0})I_{z}} \\ \nonumber
&=&\Sigma_{n}A_{n}e^{-i(\delta_{0}+n\omega_{r})(t-t_{0})I_{z}} \\ \nonumber
&=&\Sigma_{n}A_{n}\left(
\begin{array}{lr}
e^{-i[\frac{n\omega_{r}}{2}(t+t_{0})+\frac{\delta_{0}t}{2}]}&0 \\
0 &e^{i[\frac{n\omega_{r}}{2}(t-3t_{0})+\frac{\delta_{0}t}{2}]}
\end{array}\;\;\right)e^{in\omega_{r}t_{0}} 
\end{eqnarray}
Where the expansion coefficients $A_{n}$ can be found to be $A_{n}=|F_{n}|^{2}$ with\cite{mehr83}
\begin{equation}
F_{n}=\frac{1}{2\pi}\int_{0}^{2\pi}d\phi e^{i[-n\phi+\frac{C_{1}}{\omega_{r}}{\rm sin}\phi
-\frac{S_{1}}{\omega_{r}}{\rm cos}\phi+\frac{C_{2}}{2\omega_{r}}{\rm 
sin}2\phi -\frac{S_{2}}{2\omega_{r}}{\rm cos}2\phi ]}
\end{equation}
Comparing Eq.(31) with Eq.(11), we have
\begin{equation}
U_{CSn}(t)=A_{n}\left(
\begin{array}{lr}
e^{-i[\frac{n\omega_{r}}{2}(t+t_{0})+\frac{\delta_{0}t}{2}]}&0 \\
0 &e^{i[\frac{n\omega_{r}}{2}(t-3t_{0})+\frac{\delta_{0}t}{2}]}
\end{array}\;\;\right) 
\end{equation}
which, when $t_{0}=0$ is chosen, is reduced to 
\begin{equation}
U_{CSn}(t)=A_{n}\left(
\begin{array}{lr}
e^{-i\frac{\delta_{0}+n\omega_{r}}{2}t}&0 \\
0 &e^{i\frac{\delta_{0}+n\omega_{r}}{2}t}  
\end{array}\;\;\right)
\end{equation}
which can, in terms of its matrix elements, be denoted in a more concise form 
as follows
\begin{equation}
U_{CSn}^{pp}(t)=A_{n}e^{-i\frac{\epsilon_{p}(\delta_{0}+n\omega_{r})t}{2}}\;\;\;p=0,1
\end{equation}
where $p=0 (1)$ corresponds to 
spin state $|0>=|+\frac{1}{2}>$ ($|1>=|-\frac{1}{2}>$) and 
$\epsilon_{p}=p-\delta_{p,0}$ with $\delta_{p,0}$ the 
Kronecker function.
\subsection{The effects of $U_{90}^{F}$ and $U_{CS}^{F}$ on pseudo pure state}
%\paragraph{}
Suppose we have prepared a pseudo pure state $|pm>$by state labeling
techniques\cite{gers97a,cory97,cory98}
(detailed in Section IV for solid state NMR quantum computing). 
The effect of the 90$^{\circ}$ RF pulse on the state is 
\begin{eqnarray}
|pm>&\rightarrow& \Sigma_{n,l}e^{-in\omega_{r}t_{p}}|n>U_{n-l}^{90x}<l|pm> \\ \nonumber
&=&\Sigma_{n}e^{-in\omega_{r}t_{p}}|n>U_{n-m}^{90x}|p>
\end{eqnarray}
Because the only non-zero component of $U_{n}^{90x}$ is $U_{0}^{90x}$
and notice that 
$|0>\stackrel{90^{\circ}_{x}}{\longrightarrow}(|0>+i|1>)/\sqrt{2}$, 
$|1>\stackrel{90^{\circ}_{x}}{\longrightarrow}(|1>+i|0>)/\sqrt{2}$, 
the above equation is simply 
\begin{equation}
|pm>\stackrel{90^{\circ}_{x}}{\longrightarrow}\frac{1}{\sqrt{2}}|m>(|p>+i|p-\epsilon_{p}>)
\end{equation}
%\paragraph{}
The effect of the chemical shift interaction on a pseudo pure state 
can be found as follows
\begin{eqnarray}
|pn>\stackrel{H^{F}_{CS}}{\longrightarrow}
U_{CS}^{F}|pn>&=&\Sigma_{k,l}e^{-ik\omega_{r}t}U_{CSk-l}|k><l|pn> \\ \nonumber 
&=&\Sigma_{k}e^{-ik\omega_{r}t}U_{CSk-n}|pk>
\end{eqnarray}
The combined effects of the RF pulse and the chemical shift interaction 
are therefore given as 
\begin{equation}
U_{CS}^{F}U_{90x}^{F}|pm>=\frac{1}{\sqrt{2}}\Sigma_{k}e^{-ik\omega_{r}t}U_{CSk-m}
(|pk>+i|p-\epsilon_{p}k>)
\end{equation}
Eqs.(37-39) are the basic equations important to
the calculations in the following sections.
\subsection{The spectral representation of the readout signal}
The signal Eq.(24) can be decomposed into 
two terms $S^{pm}_{+}(t)=S^{pm}_{x}(t)+iS^{pm}_{y}(t)$ where 
$S^{pm}_{x}(t), S^{pm}_{y}(t)$ are obtained from Eq.(24) by replacing 
${\tilde{I_{+}}}$ with ${\tilde{I_{x}}}$ and ${\tilde{I_{y}}}$, respectively.
We will give the results here (the derivation of $S^{pm}_{y}(t)$ is 
shown in Appendix A and it equally 
applies to the calculation of $S^{pm}_{x}(t)$).
\begin{equation}
S^{pm}_{y}(t)=i\frac{1}{2}\Sigma_{k}A_{k-m}\epsilon_{p}
[e^{-i\epsilon_{p}(k-m)
\omega_{r}t}-e^{i\epsilon_{p}(k-m)\omega_{r}t}]
\end{equation}
whose spectrum is obtained by Fourier transformation 
\begin{equation}
I^{pm}_{y}(\omega)=i\frac{1}{2}\Sigma_{k}A_{k-m}\epsilon_{p}
[\delta(\omega-\epsilon_{p}(k-m)
\omega_{r})-\delta(\omega+\epsilon_{p}(k-m)\omega_{r})
\end{equation}
For a given system, the largest mode number is fixed, say, $K$,
which is restricted by the sensitivity limit.
Then the above equation means that the sideband manifold consists of
the following bands: 
$[-K-m, -K-m+1,...,K-m]\epsilon_{p}$ and 
$[-K+m, -K+m+1,...,K+m]\epsilon_{p}$.
The amplitude of each band depends on the value of $p$:
for different $p$, there is a 180$^{\circ}$ phase factor difference.
Therefore, a unique one-to-one correspondence is established 
between a pseudo pure state and a spectral representation.
%\paragraph{}
The signal $S^{pm}_{y}(t)$ (or $I^{pm}_{y}(\omega)$)
contains two groups of bands with opposite signs in intensity.
If quadrature detection is used, the signal is found to be 
\begin{equation}
S^{pm}_{+}(t)=\Sigma_{k}A_{k-m}\epsilon_{p}
e^{-i\epsilon_{p}(k-m)\omega_{r}t}
\end{equation}
and its spectrum is readily found to be
\begin{equation}
I^{pm}_{y}(\omega)=\Sigma_{k}A_{k-m}\epsilon_{p}
\delta(\omega-\epsilon_{p}(k-m)\omega_{r})
\end{equation}
which contains only one group of sidebands given by indices 
$[-K+m,-K+m+1,...,K+m]\epsilon_{p}$. 
%\paragraph{}
As a demonstration, we choose the six-level system shown in 
Fig.2 in which $|K|=1$. Then the
spectral representations of all the pseudo pure states 
are given in Fig.3 and Fig.4 for single crystal and powder
samples, respectively.
The one-to-one correspondence between state and spectrum is obvious.
The pseudo states are differentiated by amplitude distribution or
by a phase factor or both. 
The spectra are absorptive for powder sample but
for single crystal, they may be dispersive 
but in the later case,
the relative phase differences can still be used to unambiguously 
identify the particular pseudo pure state of the solid state NMR quantum computer  
before the readout RF pulse was applied. 

\section{State labeling}

\subsection{General}

State labeling is the unique feature of ensemble quantum computing
because a pure state is not naturally available in an ensemble. 
To carry out quantum computation, one must first "purify" the 
ensemble so that it can be regarded as being in a pure state. 
The problem of state labeling in solution NMR has been well 
addressed. There are two types of labeling: one is temporal and
another spatial. The former uses proper time averaging
and the latter uses spatial averaging, of the density matrices, to 
construct a pseudo pure state. Here we show how these methods
can be extended to Floquet space quantum computing. 

Given an initial density matrix, $\rho^{F}(t_{0})$, state labeling 
renders it a pseudo pure state and a quantum computation task  
can be undertaken starting from it as follows
\begin{equation}
P^{F}U^{F}\rho^{F}(t_{0}){U^{F}}^{-1}=C|\phi_{0}^{F}><\phi_{0}^{F}|=c\psi^{F}
\end{equation}
where $C$ is the computation operator and
$c$ is a constant coefficient and $P^{F}$ a certain 
preparation (not unitary in most cases). 

State labeling from the initial (thermal state) density matrix 
given by $<n|\sigma^{F}(0)|m>=\sigma(0)\delta_{0,m}\delta_{0,n}$ 
is a trivial task because there are only two non-zero terms for spin-1/2
systems. Either with phase cycling (two experiments) or by
applying a gradient field,
a specified state can be chosen using the routine techniques.
However, this is not the general case because, first, the initial density 
matrix in formalized Floquet theory can be chosen as a different form 
$<n|\sigma^{F}(0)|m>=\sigma(0)\delta_{m,n}$ and secondly, state labeling 
may need to start with a density matrix other than the thermal equilibrium 
form. In addition, certain computations may require a pure state 
with the mode indices not equal to zero. 
Therefore, in the following two general techniques will be given
which can be employed for any type of mixed state.
 
\subsection{Multi-pulse techniques}
From Section III, a pure state corresponds to certain peak profile in
an MAS spectrum. The preparation of a pseudo pure state, therefore, 
is equivalent to constructing a subspectrum with specific peak profile.
Over the past decades, there have been developed an array of 
techniques in solid state NMR to manipulate the MAS spectral peaks,
such as TOSS (total suppression of sidebands), PASS(phase adjusted 
spinning sidebands), IRS(isotropic rotation sequence) and their 
combinations and two-dimensional extensions etc\cite{antz99}. 
As an example, we demonstrate that the 2D-PASS sequence is 
a satisfactory technique for state labeling. The sequence 
consists of five $\pi$ pulses and
is shown in 
Fig. 5. The parameter $\Theta$ (pitch) is used to characterize the
separations between each adjacent pair of $\pi$ pulses.
In 2D-PASS experiment\cite{antz95}, $\Theta$ also represents the first dimension
and it is changed systematically according to prescribed 
separations between pulse pairs such as
\begin{eqnarray}
2\Sigma_{q=1}^{n}(-1)^{q}e^{im\theta_{q}}+1-(-1)^{n}e^{im(\Theta+\theta_{T})}
&=&0 \\ \nonumber
-2\Sigma_{q=1}^{n}(-1)^{n+q}\theta_{q}+\theta_{T}=0 \\ \nonumber
\theta_{T}&=&0 
\end{eqnarray}
where $n$ is the total number of the sidebands in the MAS spectrum.
The meaning of other parameters is shown in Fig.5.
Each sideband (or central band) can be extracted from the projection 
along the first dimension. When necessary, the intensity of each band 
can be modified by adding a pair of pulse $90^{\circ}_{-x}\theta_{x}$
at the end of the PASS sequence, where the value of $\theta$ is determined
by the desired intensity of the band. The peak profile of a
pseudo pure state can then be constructed by adjusting $\Theta$
and $\theta_{x}$ systematically.
For example, given a peak
profile ${A_{K}, K=1,2,...,n}$ where $A_{K}$ is the intensity of
$K$-$th$ sidebands. The second dimension signal (FID) in a 2D-PASS 
experiment is given by 
\begin{equation}
S(t_{2})=\Sigma_{k=-\infty}^{\infty}a^{(k)}e^{-ik\Theta}
e^{i(\delta_{0}+k\omega_{r})t_{2}}
\end{equation}
where the explicit expression of $a^{(k)}$ can be found in Ref.[77].
The simplest peak "basis" can be chosen as
\begin{equation}
\Sigma_{\Theta}S^{\Theta}(t_{2})x_{K}^{\Theta}=
A_{K}e^{i(\delta_{0}+K\omega_{r})t_{2}}
\end{equation}
where ${x_{K}}$ ($=sin\theta_{x}$) are to be determined.
By summing all possible values of $\Theta$ the 
above equation can be simplified as
\begin{equation}
\Sigma_{\Theta}a^{(k)}e^{ik\Theta}x_{K}^{\Theta}=\delta_{kK}A_{k}
\end{equation}
Therefore, the values of ${x_{K}}$ can be found out by solving 
the above linear equations 
\begin{equation}
x=A^{-1}\Delta
\end{equation}
where the matrices $A={a^{(k)}e^{ik\Theta}}$ and $\Delta={\delta_{kK}A_{K}}$.
It can be easily verified that $A^{-1}$ exists.

Therefore, any subspectral
profile can be constructed with this sequence: 
$90_{x}^{\circ}-[ASL]-90_{x}^{\circ}-\theta_{-x}$.
The number of the values of $\Theta$ is determined by
the number of the sidebands in the MAS spectrum and in
principle there are no restrictions on the values of $\Theta$.
Therefore, this technique can readily deal with a system with
as many as hundreds or even thousands of sidebands.

To reduce the number of steps to a practically acceptable level, 
the density matrix may be simplified before the standard state labeling 
techniques is invoked. We believe that the selective excitation
techniques such as Dante pulse sequence\cite{erns87}, can be incorporated in 
the preparation of the density matrix for state labeling.

\subsection{Gradient field selection}

Temporal labeling methods as discussed above are easy to implement and 
retain the sensitivity of the whole sample, but  
usually require a number of experiments. This reduces
computing efficiency and in the least-promising cases might conceal
the advantage
of quantum
parallelism\cite{gers97a,chua98a}. Therefore, temporal labeling is best
suitable for low-qubit 
implementations. Another technique, {\it i.e.}, spatial labeling was proposed
by Cory {\it et al.}\cite{cory97,cory98} which may or may not use phase cycling. 
Here this method is extended to the solid state MAS NMR case. 

In a magnetic gradient field, the Floquet energy levels are
dispersed along the direction of gradient. The effect 
of a gradient field on the energy 
levels is the same as in normal Hilbert space. 
Without losing generality, we consider the case where the gradient field is 
along the z-direction (parallel with the static magnetic field).
We use the "sandwich" pulse sequence $G_{z1}(t_{G1})-90^{\circ}_{x}-G_{z2}(t_{G2})$
where $t_{G1}, t_{G2}$ are the gradient pulse lengths and $90^{\circ}$ 
is the RF pulse, as shown in Fig. 6. The evolution 
operator during the gradient pulse can be from Eq.(11) 
\begin{eqnarray}
U_{G}(t,t_{0})&=& e^{-izG_{z}\int_{t_{0}}^{t}dt(\omega_{CS}(t)+\delta_{0})I_{z}} \\ \nonumber
&=&\Sigma_{n}A_{n}\left(
\begin{array}{lr}
e^{-izG_{z}[\frac{n\omega_{r}}{2}(t+t_{0})+\frac{\delta_{0}t}{2}]}&0 \\
0 &e^{izG_{z}[\frac{n\omega_{r}}{2}(t-3t_{0})+\frac{\delta_{0}t}{2}]}
\end{array}\;\;\right)e^{in\omega_{r}t_{0}} \\ \nonumber
&=&\Sigma_{n}U_{G\;n}(t)e^{in\omega_{r}t_{0}}
\end{eqnarray}
from which the evolution operator in Floquet space can be 
easily found using Eq.(10).

With any given initial state $\rho^{F}(t_{0})$, the 
final density matrix after the sandwich pulse sequence is given by
\begin{equation}
\rho^{F}(t)=U_{G2}^{F}(t_{2},t_{0}^{'})U_{90x}^{F}U_{G1}^{F}(t_{G1},t_{0})
\rho^{F}(t_{0})
{U_{G1}^{F}}^{-1}(t_{G1},t_{0}){U_{90x}^{F}}^{-1}{U_{G2}^{F}}^{-1}(t_{2},t_{0}^{'}) 
\end{equation}
where $t_{2}=t_{G2}+t_{0}^{'}, t_{0}^{'}=t_{0}+t_{G1}+t_{p}$.
It is easy to show that above equation only provides a constraint on the
the Floquet space if the gradient field is time 
independent. 
Using Eq.(10), we can find the condition for state labeling from Eq.(51) as
\begin{equation}
\epsilon_{p}k\omega_{r}G_{z2}t_{G2}+\epsilon_{q}l\omega_{r}G_{z1}t_{G1}=0 
\end{equation}
where $k, l$ are integers.
The detailed derivation of above equations is given in Appendix B. 
By setting $G_{z1}, G_{z2}, t_{G1}$ and $t_{G2}$ properly, the 
desired situation
that only one term survives the pulse sequence can be realized,
thus a pseudo pure state is prepared.

Obviously, the gradient field method can also be incorporated with 
selective excitation but its merit is unknown. 
In the following, we will focus on the techniques that do not 
need a gradient. The thorough investigation of the application 
of a gradient field in solid state NMR quantum computing will be
presented elsewhere.

\section{Universality and Gate Design}

The above procedures for implementing state labeling can be 
extended to
construct basic operating matrices, {\it i.e.}, elementary 
quantum logic gates. 
The universality can be ensured if all possible operations 
can be realized with a set of elementary gates.
The central problem is how to realize the operations
that transform any state into any other state.
As an example, we show that the 2D-PASS sequence can be used
as a basic pulse block in constructing universal logic gates.
From "peak manipulation" point of view, if all peak profiles
can be realized starting from any peak profile, the gates
thus constructed are universal. \\

Here we give a general discussion on the relationship between
peak manipulation and the construction of 
complete unitary transform (universal gates).
Starting from the initial density matrix $\rho^{F}(0)$,
two pseudo pure states can be prepared as follows 
\begin{eqnarray}
\rho^{F}_{1}&=&P_{1}^{F}\rho^{F}(0) \\ \nonumber
\rho^{F}_{2}&=&P_{2}^{F}\rho^{F}(0)
\end{eqnarray}
Thus the unitary transform that maps state $\rho_{1}^{F}$
to state $\rho_{2}^{F}$ can be realized with
\begin{equation}
U_{12}^{F}=P^{F}_{2}{P_{1}^{F}}^{-1}
\end{equation}
Suppose there are $L$ Floquet states that are usable for
quantum computing where $L=L_{d}+L_{u}$ where $L_{d}, L_{u}$
are the number of states corresponding to spin 
quantum number $\frac{1}{2}, -\frac{1}{2}$, respectively.
Its is easy to find from the energy levels that the total 
number of peaks
from this energy level manifold is $L-1$. 
Therefore, for a unitary matrix of $M\times M$ dimension,
$M+1$ Floquet states are needed. 
The unitary matrices can then be determined
from the peak manipulation matrices which constitute
a complete set:
\begin{equation}
U_{mn}^{F}=\Sigma_{i,j}a_{ij}^{mn}P_{ij}^{mn}
\end{equation}
The unitary matrices thus constructed necessarily form
a complete set because peak manipulation matrices are complete.

The experimental realization of the universal operation can be
implemented with, {\it e.g.}, 2D-PASS sequence [ASL]
shown in Fig.5. 
For example, pulse sequence $\theta_{x}^{(n)}-90^{\circ}_{-x}
-[ASL]^{-1}-90^{\circ}_{-x}-90^{\circ}_{x}-[ASL]-90^{\circ}_{x}
-\theta^{(m)}_{-x}$, where $[ASL]^{-1}$ is the time-inverted 
version of the ASL sequence shown in Fig.5,
gives the pulse sequence that transforms a
pseudo pure state $|pm><pm|$ to another $|qn><qn|$. 
With these transform matrices, therefore, any possible
operations compatible with the system are realizable.

This method, perhaps rather forceful, is workable in practice, 
at least for low qubit cases. However,
we point out that there may exist more efficient pulse 
sequences that can realize above transformations
and that is to be the major goal of our subsequent
efforts.
\section{Example: Grover's search}
In this section, we demonstrate the use of the
theoretical formalism developed in the preceding sections 
to implement experimentally a quantum algorithm.
We will use Grover's search \cite{grov97} as an example. 

Grover's search consists of four steps\cite{grov97}: (1)the 
preparation of pseudo pure state; (2) HW transform;
(3) conditional flipping; (4) average about the mean;
where step (3) and (4) are repeated $2\sqrt{N}/\pi$
times for an $N-$item search. 
When using the 2D-PASS sequence,
steps (1-3) can be incorporated in one experimental step
with the initial density matrix as the input and
an equi-amplitude superposed state as the output. 
In the output state, the phase difference between 
the flipped bit and the rest is 180$^{\circ}$.   
The step (4) corresponds to the transform matrix 
with elements $U_{ij}=\frac{1}{2}-\delta_{ij}$,
each of which can be implemented according to Eq.(55).
Generally, the corresponding pulse sequence can then 
be designed as follows: 
$90^{\circ}_{x}-[ASL]-90^{\circ}_{x}-\theta^{(l)}_{-x}-
\theta_{x}^{(m)}-90^{\circ}_{-x}-[ASL]^{-1}-
90^{\circ}_{x}-[ASL]-90^{\circ}_{x}-\theta^{(n)}_{-x}-Acquisition$
with $l, m, n=1,2,3,4...$ 
This means the number of experiments is $l\times m\times n$
which would soon become impractically large. However,
above construction of the pulse sequence may reduce the
number of experiments substantially. For example,   
for the 2-qubit search experiment demonstrated 
in this work, however, only four experiments are required
for each search.
The Floquet energy level diagram and four "working" states
are illustrated in Fig. 7 (A) and (B), respectively.

The experiments were performed on a Bruker MSL-200 NMR
spectrometer. The $^{13}$C resonance frequency was
50.3 MHz. The sample was hexamethylbenzene (HMB) 
which was spun at the magic-angle with a spinning speed
of 5 kHz. The aromatic carbon atom which has a chemical shift anisotropy 
of about 100 ppm is the spin we used for
quantum computing. The methyl carbon has a small chemical shift anisotropy
and its peak is used as the phase standard of the spectra.
With the spinning speed used, there are two sidebands
that are clearly observable for the aromatic carbon. 
Based on pulse sequences and the procedures discussed in 
the previous section, four search results were obtained and
are shown in Fig. 8. As seen from Fig.7(B) and Fig.8,
the theoretical prediction is in good agreement with the
experimental result. 

\section{Conclusions}
In this work, magic-angle spinning solid state NMR QC theory is 
developed based on 
formalized Floquet theory.
Because the mode space is controlled by the sample spinning speed,
the realizable number of qubits is changeable and can be made as 
large as the task demands (the ultimate limitation comes from 
signal to noise ratio). 
The techniques required for state manipulations
(labeling, coherence 
selection etc) 
are analyzed. The spectral representations of state is given that is 
crucial to readout of QC registers. The basic QC gates are demonstrated 
and an important QC algorithm is shown to be realizable in solid state 
quantum computer. Based on these theoretical analyses, it has become 
clear that solid state NMR may become a good alternative methodology for ensemble
quantum computing.
This technque has many advantages: such as large working space
(Floquet) that is adjustable. 
This makes a vast difference between this
technique and previously proposed NMR-related methods because the
dimension of the computing space can be augmented significantly for
the spin systems which otherwise can only offer very low number of qubits.
The sensitivity is generally higher than its liquid state 
counterpart of the same size because of the difference in the number 
densities of nuclear spins. The pulse sequences are conveniently implemented 
with conventional solid state NMR techniques although non-conventional 
techniques may be used to enhance sensitivity and achieve higher-qubit operations.
We point out that the number of qubits currently manageable in our initial experiment is rather limited mainly because of the brute force pulse sequences used. We believe, however, more efficient pulse sequences for implementing higher-qubit QC operations in Floquet space can be found.

The chemical shift interaction of single spin-1/2 system is 
exemplified in this work, but the principles and 
procedures used here can be extended to other interactions 
such as dipolar and quadrupolar interactions, which is the object of 
further work. In fact, when more spins are involved in coupled systems
or large quadrupolar interaction is present, high qubits are more conveniently 
realized in those systems. Nevertheless, we recognize that the difficulty involved in manipulating experimentally these systems may be significantly greater.

Finally, for a practical quantum computer, error correction is necessary. The recently published error correction schemes must be modified to accommodate solid state NMR based quantum computing in Floquet space.
\acknowledgments
{This work was supported by the Hundred Talents Program of the 
Chinese Academy of Sciences and the Natural Science and Engineering Research Council
of Canada. SD acknowledges the stimulating discussions with Prof. Guo Wei 
Wei of the National University of Singapore. He also thanks Dr. Ole N. 
Antzutkin of the Laboratory of Chemical Physics of the National Institute of 
Diabetes and Digestive and Kidney Diseases, NIH (USA) for sending a 
reprint of his review paper\cite{antz99} on sideband manipulation techniques 
that are important in experimental implementation of solid state NMR 
quantum computing.}

\appendix {\bf Appendix A: Derivation of Eq.(40) and Eq.(41)} \\

%\paragraph{}
From Eqs.(24,39), the signal $S^{pm}_{y}(t)$ can be written as
\begin{eqnarray}
S^{pm}_{y}(t)&=&\frac{1}{2}{\rm
Tr}\{\Sigma_{k,k'}e^{-ik\omega_{r}t}U_{CSk-m}(|pk>+|p-\epsilon_{p}k>)
e^{ik'\omega_{r}t}(<pk'|+|p-<\epsilon_{p}k'|) \\ \nonumber
& &U_{CSk'-m}^{-1}\Sigma_{l,j}|l>I_{y}<l+j|e^{ij\omega_{r}t}\} \\ \nonumber
&=&\frac{1}{2}{\rm Tr}\{\Sigma_{k,k',j,l}e^{-i(k-k'-j)\omega_{r}t}
U_{CSk-m}(|pk>+|p-\epsilon_{p}k>)
(<pk'|+<p-\epsilon_{p}k'|) \\ \nonumber
& &U_{CSk'-m}^{-1}|l>I_{y}<l+j|\}
\end{eqnarray}
In terms of matrix elements, it becomes 
\begin{eqnarray}
S^{pm}_{y}(t)&=&\frac{1}{2}\Sigma_{k,k'j,l}\Sigma_{s,n} 
e^{-i(k-k'-j)\omega_{r}t}(U_{CS k-m}^{sp}+U_{CS k-m}^{s p-\epsilon_{p}})<n|k> \\ \nonumber
& &\Sigma_{q,n'}({U_{CSk'-m}^{pq}}^{-1}
+{U_{CSk'-m}^{p-\epsilon_{p}\;q}}^{-1})<n'|k'> 
<n'|l><q|I_{y}|s><l+j|n> \\ \nonumber 
&=&\frac{1}{2}\Sigma_{k,k'}\Sigma_{s,q}[U_{CSk-m}^{sp}+U_{CSk-m}^{s p-\epsilon_{p}}]
[{U_{CSk'-m}^{pq}}^{-1}+{U_{CSk'-m}^{p-\epsilon_{p} q}}^{-1}]I_{y}^{qs}
\\
\nonumber
&=&\frac{1}{2}\Sigma_{k,k'}\Sigma_{s,q}[U_{CSk-m}^{sp}{U_{CSk'-m}^{pq}}^{-1}
+U_{CSk-m}^{sp}{U_{CSk'-m}^{p-\epsilon_{p} q}}^{-1}
+U_{CSk-m}^{s p-\epsilon_{p}}{U_{CSk'-m}^{pq}}^{-1}
+U_{CSk-m}^{s p-\epsilon_{p}}{U_{CSk'-m}^{p-\epsilon_{p} q}}^{-1}] 
\end{eqnarray}
Note that $U_{CSn}$ is diagonal and $I_{y}$ is Hermitian 
and its diagonal elements are all zero. The above equation only has two non-zero 
terms given as
\begin{equation}
S^{pm}_{y}(t)=\frac{1}{2}\Sigma_{k,k'}
[U_{CSk-m}^{pp}{U_{CSk'-m}^{p-\epsilon_{p}\;p-\epsilon_{p}}}^{-1}I_{y}^{p-\epsilon_{p}\;p}
+U_{CSk'-m}^{p-\epsilon_{p}\;p-\epsilon_{p}}{U_{CSk-m}^{pp}}^{-1}I_{y}^{p\;p-\epsilon_{p}}]
\end{equation}
Using the matrix expression of the $U_{CSn}(t)$ Eq.(34) 
and the orthogonal properties of Bessel functions contained 
in coefficients $A_{n}$,
the time-domain signal Eq.(40) is readily obtained. 
%\paragraph{}
The calculation of $S^{pm}_{x}(t)$ is completely the same as given above
and its result is 
\begin{equation}
S^{pm}_{x}(t)=\frac{1}{2}\Sigma_{k}A_{k-m}\epsilon_{p}
[e^{-i\epsilon_{p}(k-m)
\omega_{r}t}+e^{i\epsilon_{p}(k-m)\omega_{r}t}]
\end{equation}
The quadrature detection signal (eq.(42)) is 
the sum of above equation and Eq.(40). 

\appendix{\bf Appendix B: Derivation of Eq.(52)} \\

The effect of a gradient magnetic field is used to select certain orders of
coherences from the initial density matrix by spatially averaging out others.
An arbitrary element of $\rho_{0}^{F}$ is denoted as $|pi><qj|$ with
coherence order of $p-q$ and $i-j$ in Hilbert and mode space,
respectively. To see how gradient field select density matrix elements, we
first calculate the following term (see Eq.(51))
\begin{equation}
U^{F}_{G_{z2}}(t_{2},t_{0}^{'})U_{90^{\circ}x}^{F}U^{F}_{G_{z1}}(t_{G1},t_{0})
|pi>|qj|
\end{equation}
which can be written as according to Eqs.(10,28)
\begin{eqnarray}
& &\Sigma_{n,m,l,k}U_{G_{2}n-m}U^{F}_{90^{\circ}x}U_{G_{1}l-k}|n><m|l><k|pi|<qj|
e^{i(n-m)\omega_{r}t} \\ \nonumber
&=&\Sigma_{n,m}U_{G_{2}n-m}U^{F}_{90^{\circ}x}U_{G_{1}m-i}|pm><qj|e^{i(n-m)
\omega_{r}t} \\ \nonumber
&=&\Sigma_{n,m}U_{G_{2}n-m}E^{-i[\omega_{1}\sigma_{n'}|n'>I_{x}<n'|
+n'\omega_{r}{\bf 1}|n'><n'|]t_{p}}U_{G_{1}m-i}
|pm><qj|e^{i(n-m)\omega_{r}t} \\ \nonumber
&=&\Sigma_{n,n',m}U_{G_{2}x}U_{I_{x}}|n'><n'|e^{in'\omega_{r}t_{p}}
U_{G_{1}m-i}|pm><qj|e^{i(n-m)\omega_{r}t} \\ \nonumber
&=&\Sigma_{n,m}U_{G_{2}n-m}U_{I_{x}}U_{G_{1}m-i}|pm><qj|e^{i\omega_{r}(mt_{p}
+(n-m)t} 
\end{eqnarray}
where $t_{p}$ is the 90 degree pulse width and $t=t_{2}-t_{0}$.
Assuming $\rho^{F}(t_{0})=|pm><qj|$, Eq.(51) is then written as 
\begin{equation}
\Sigma_{n,m,n'm'}U_{G_{2}n-m}U_{I_{x}}U_{G_{1}m-i}|pm><qm'|U^{-1}_{G_{1}m'-j}
U^{-1}_{I_{x}}U^{-1}_{G_{2}n'-m'}e^{i\omega_{r}[(m-m')t_{p}
+(n-n'+m'-m)t]} 
\end{equation}
Let $t_{0}=0$. Using Eq.(50), the terms related to the gradient field
in above equation are of the form 
\begin{eqnarray}
& &e^{\pm i\frac{n-m}{2}G_{1z}z\omega_{r}t}, 
e^{\pm i\frac{m-i}{2}G_{1z}z\omega_{r}t} \\ \nonumber
& & e^{\pm i\frac{n'-m'}{2}G_{2z}z\omega_{r}t},
e^{\pm i\frac{m'-j}{2}G_{1z}z\omega_{r}t}
\end{eqnarray}
Note that the gradient fields are symmetrical with respect to
$z$ axis and the indices $n,m,n',m'$ run from $-\infty$ to $\infty$.
Therefore, above equation would vanish unless
\begin{equation}
(i-j)G_{1}t_{G_{1}}=(n-n')G_{2}t_{G_{2}}
\end{equation}
If we denote the coherence orders in the mode space during the first
and second gradient field pulses, $i-j$ and $n-n'$, as $l$ and $k$,
respectively, above condition is the same as Eq.(52) by adding the
coherence transfer condition in Hilbert space.

\figure {{\bf Fig. 1}
(a) The Floquet energy levels of spin-1/2 under MAS.
(b) The read pulse (90$^{\circ}$) for a Floquet state. }
\figure {{\bf Fig. 2}
Upper: the sub-manifold of six energy levels of a spin-1/2 system under MAS
and lower: the representation of each state.}
\figure {{\bf Fig. 3}
The readouts of the Floquet system shown in Fig.2 
for a single crystal sample. The parameters are $\delta_{CS}$=20 kHz, $\eta_{CS}$
=0.5, the spinning speed $\omega_{r}$=4 kHz. The relative orientation of the
crystal with respect to the magnetic field is described by two Euler angles
(between the lab frame and the principal axis system of the CSA tensor) 
($\alpha, \beta$)=($30^{\circ}, 60^{\circ}$).   
}
\figure {{\bf Fig. 4}
The readouts of the Floquet system shown in Fig.2 for
real polycrystalline powder sample. The 
parameters: $\delta_{CS}$=20 kHz, $\eta_{CS}$=0.5 and 
spinning speed $\omega_{r}$=4 kHz.
}
\figure {{\bf Fig. 5}
The pulse sequence of a 2D PASS experiment for spin-1/2 systems
proposed by Atzutkin, Shekar and Levitt\cite{antz95}[ASL].
}
\figure {{\bf Fig. 6}
The typical pulse arrangement for state labeling with gradient field selection.
}
\figure {{\bf Fig. 7}
Grover search based on MAS NMR. 
(a) Energy level diagram showing the states used for the experiment.
(b) Schematic representation of the 
four Floquet states chosen for the experiment.  
}

\figure {{\bf Fig. 8}
The experimental result 
of four possible states with Grover search algorithm 
based on MAS NMR,
represented with the sideband pattern of the
aromatic carbons of hexamethylbenzene. The rightmost peak comes
from the methyl carbons and is used as phasing reference.}\\

\end{document}